# Cyberspace security: How to develop a security strategy

**Bel G. RAGGAD**, PhD
Seidenberg School of CS & IS
Pace University, Pleasantville,NY
10570
braggad@pace.edu

**Sahbi SIDHOM**, PhD
LORIA & University of Nancy,
BP. 239, 54506 Vandoeuvre cedex -
France
Sahbi.Sidhom@loria.fr

**Abstract:**

Despite all visible dividers, the Internet is getting us closer and closer, but with a great price. Our security is the price. The international community is fully aware of the urgent need to secure the cyberspace as you see the multiplication of security standards and national schemes interpreting them beyond borders: ISO 15408, ISO 17799, and ISO 27001.

Even though some countries, including the Security Big Six (SB6), are equipped with their security books and may feel relatively safe; this remains a wrong sense of security as long as they share their networks with entities of less security.

The standards impose security best practices and system specifications for the development of information security management systems. Partners beyond borders have to be secure as this is only possible if all entities connected to the partnership remain secure. Unfortunately, there is no way to verify the continuous security of partners without periodic security auditing and certification, and members who do not comply should be barred from the partnership. This concept also applies to the cyber space or the electronic society. In order to clean our society from cyber crimes and cyber terrorism we need to impose strict security policies and enforce them in a cooperative manner.

The paper discusses a country's effort in the development of a national security strategy given its security economic intelligence position, its security readiness, and its adverse exposure.

**Key words:**

Security strategy, security auditing, Security economic intelligence, Security Big Six (SB6), security metric, cyberspace, information security management, system specifications, certification.

## 1. Introduction

The Internet is getting us closer and closer, but with what price. The international community is fully aware of the urgent need to secure the cyberspace (White House, 2003) as you see the multiplication of security standards and national schemes interpreting them beyond borders: ISO 15408, ISO 17799, and ISO 27001 (GuideInformatique, 2007).

Even though some countries, including the Security Big Six (SB6), are equipped with their security books and may feel relatively safe; this remains a wrong sense of security as long as they share their networks with entities of less security.

The standards impose security best practices and system specifications for the development of information security management systems. Partners beyond borders have to be secure as this is only possible if all entities connected to the partnership remain secure. Unfortunately, there is no way to verify the continuous security of partners without periodic security auditing and certification, and members who do not comply should be barred from the partnership. This concept also applies to the cyber space or the electronic society. In order to clean our society from cyber crimes and cyber terrorism we need to impose strict security policies and enforce them in a cooperative manner.

The paper discusses a country's effort of the development of a national security strategy (White House, 2006) given its security economic position, its security readiness, and its adverse exposure.

## 2. Security economic intelligence

Can a developed country get improved access to information or enjoy the continuous development of technologies and networks that industrial countries have (Kekic, 2007)? We thought global computing is the key feature of future peace and prosperity for all. All need a diversified net of resources where information feasibly fused to provide decision support, and hence the quality of life for all. There is need for applications to support decision-making by organizations, enterprises or individuals; and because individuals, are connected to organizations, at the input, process, or output, organizations are also connected to society and the homeland.

This will facilitate better and, eventually, more innovative approaches and answers to the opportunities and risks in a rapidly changing world. Economic Intelligence (EI) aims to take advantage of this opportunity to develop better methods for the identification of relevant sources of information, the analysis of the collected information and its manipulation to provide what the user needs for decision making (LOYOLA, 2007), (SITE, 2006).

This study intends to emphasize a special version of EI, referred to here as the security economic intelligence. Security Economic Intelligence (SEI) mainly addresses organizations and countries that crave for up-to-date security information to make the best security decisions, in terms of disaster recovery planning, business continuity, and homeland security, in a framework of a defined security strategy. Unfortunately, while this concept is very well defined for all SB6 countries, the rest of the world is still lagging behind and have no feasible way to know more or even to assemble the resources to process the security knowledge they seek and act upon it. Reasons for failure on the SEI concept in non-SB6 countries include:

- Lack of sharing for useful security knowledge
- Imposed security standards
- Imposed security certifications
- Non-exportability of useful security solutions
- Infeasibility of acting uninformed

While global computing and Internet power are made to share useful knowledge, it seems awkward that the most useful knowledge of all, the security knowledge that without it these technologies will be of no use, cannot be shared by all. You will see, instead, endless security standards and certifications that have become barriers for those who cannot afford it, like many of the non-SB6 countries to lag behind.

Moreover, when security economic intelligence is absent, the SB6 countries would not export their useful security solutions and the would not share their security economic intelligence to non-SB6 countries, may be to few friends. With the absence of security economic intelligence, however, the only remaining alternative is to act uninformed. Unfortunately uninformed security decisions may do more harm than good.

While economic intelligence relate more to foreign economic resources, activities, and policies including the production, distribution, and consumption of goods and services, labor, finance, taxation, commerce, trade, and other aspects of the international economic system, we mainly target that intelligence that cope with security resources in terms of data, information, and knowledge, activities, technology, and infrastructure needed to provide for the security of the country. The security intelligence (CSARS, 2006) sought is useful to develop and implementable security strategy that set strategic directives and strategic, functional, and operational plans for the following main goals:

-business continuity
-disaster recovery
-homeland security

*"Strategy is the great work of the organization. In situations of life or death, it is the Tao of survival or extinction. Its study cannot be neglected." (*SUN TZU (By LIONEL GILES, M.A.), 1910).

We do not intend to study how use the security economic intelligence in developing a security strategy for a nation or its organizations, as this is beyond the scope of this paper, we emphasize the fact that this new concept is at the core of any national security strategy, and it also remains as a sequential supportive process for all phases of the security strategy, as show in Figure 1.

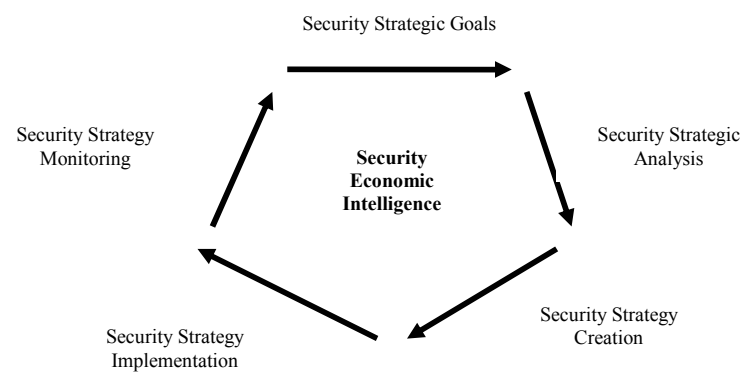

Figure 1: Phases for the development of a national security strategy

### 3. Imposed Infeasibility of security

Spend all the time there is and all the money there is, and you still cannot digest the security literature that may immunize you from insecurity; Spend all the time there is and all the money there is, and you still cannot explore the never-ending list of standards to conform to save your core from insecurity; Spend all the time there is and all the money there is, and you still cannot span the endless collection of security solutions you can afford to live with no worry of insecurity; And spend all the time there is and all the money there is, and you still cannot afford complying with the security certifications (Schneier, 2007) you need to satisfy partners' requirements, government requirements, and local, regional, and international standard.

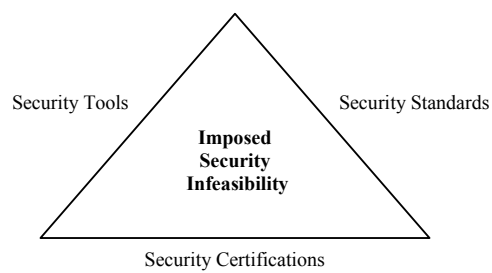

Figure 2: Imposed infeasibility of security

Due to all the imposed infeasibility constraints we explained earlier, as show in Figure 2, it will be very difficult for a non-SB6 country to afford devising implementable strategies to protect its infrastructures, businesses, and national security. There are simply too many security standards (infosyssec, 2007) to conform to, too many security certifications to comply with, and obviously too many security controls to need for adequate security.

In addition to the imposed infeasibility constraints, the ultimate advantage goal set to global computing and the sharing of computing and information resources cross-borders remains decelerated by the new cyberspace laws that are slowly populating but start refraining the still slow flow of productive Internet. Of course, the bad Internet (use for crimes, terrorism, porn, and so on) is already beyond our sense of command.

Let us delineate our insecurity and depict a simple stance for our obvious security needs, something that apply to all, including an non-SB6 country. Figure 3 illustrate this state of scrutiny.

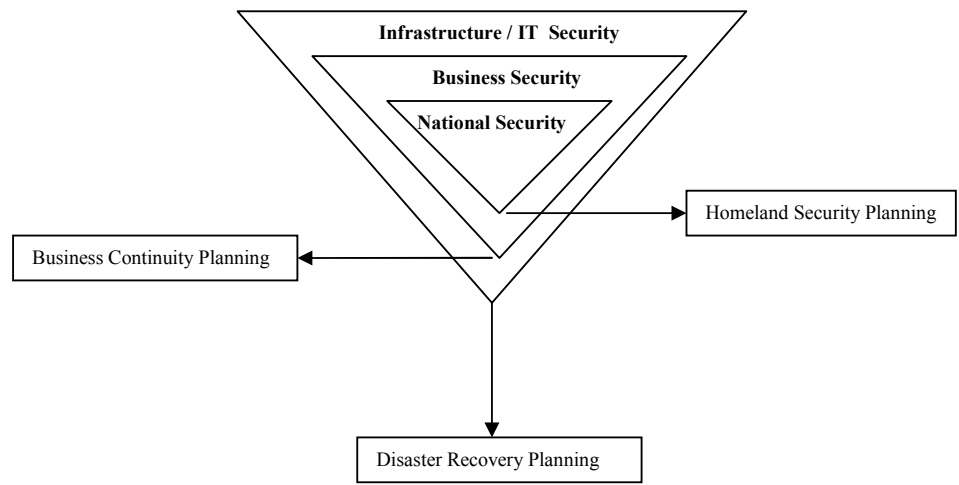

Figure 3: Implementable security strategic components

## 4. Business Continuity Planning

Business continuity is the ability of an organization to respond to disaster or business disruption through the timely detection of the disruption event, the accurate measurement of risks and business losses, and the efficient resumption of business operations.

The National Fire Protection Association (NFPA), defined business continuity as follows (NFPA-1600, 2007):

> *"Business Continuity is an ongoing process supported by senior management and funded to ensure that the necessary steps are taken to identify the impact of potential losses, maintain viable recovery strategies, recovery plans, and continuity of services."* (NFPA 1600, 2007).

This definition requires that management has to fund and support the business continuity effort. Business continuity is an incessant process that keeps track of all possible losses and their impacts on the organization. It is also responsible of maintaining any viable safeguards capable of an effective and quick recovery and continuity of business services.

### 4.1. Disaster recovery planning

Disaster recovery is the activity of resuming computing operations after a disaster, like floods, severe storms, or geologic incidents, takes place. Restoring the computing environment is often achieved through the duplication of computing operations. Disaster recovery is also concerned with routine off-site backup that IT functions migrate to in case a disaster occurs. as well as a procedures for activating vital information systems (Pastor and Cunha, 2005) in a safer computing environment.

The quality of disaster recovery is given by the organization's ability to recover information systems quickly after a disaster. While a good disaster recovery system allows the organization to have its computing operations running immediately after the occurrence of the disaster, with a poor disaster recovery system, the organization may take several days, or for ever, to restore its computing environment.

The scope of the design of a disaster recovery plan depends on many factors including the type of the incident, data affected, and business losses. The recovery from a disaster attempts to re-establish a computing environment configuration that can produce acceptable business operations' conditions.

The World Trade Center in New York was attacked on September 11, 2001, and you cannot imagine how many businesses in NY, in the USA, and in the world, were affected. Recoveries from such a disaster have been slow and very costly. Some businesses have never recovered from this disaster until today. The literature (NFPA 1600, 2007) reported that the cost of business interruptions for the World Trade Center attacks reached between $35 billion and $70 billion.

The undesired incidents may be as simple as a faulty connection or a bigger incident like a natural disaster that may be a man-made incident like fire, or an act-of-God like an earthquake or tornado. In case of a big fire that burns the infrastructure, a local redundancy system may not be effective, and an immediate continuation of business will not be possible. Instead, there will be need to have a disaster recovery plan capable of rebuilding the necessary components and installing them for the purpose of bringing back any acceptable conditions for business operations.

An organization's computing environment is becoming larger, more complex, and even more integrated. It is very rare that a piece of information, at an information resource, anywhere in the computing environment, is compromised without affecting other areas of the computing environment (ARC, 1992). Direct financial losses in one area of the computing environment can rapidly propagate from one functional unit to another functional unit. Indirect financial losses can travel from one unit to another through the loss of malcontent customers and their complaints.

### 4.2. Homeland security

Homeland security refers to the general national effort by federal, state, local agencies to protect the territory of the United States from hazards both internal and external, natural and man-made. Homeland security is officially defined by the National Strategy for Homeland Security (White House, 2002) as:

> *"a concerted national effort to prevent terrorist attacks within the United States, reduce America's vulnerability to terrorism, and minimize the damage and recover from attacks that do occur."* White House.

While it had been used only in limited policy circles before, the term has now gained more use in the United States following the September 11, 2001 disaster. While it is probably more concerned with Emergency and preparedness and response, for both terrorism and natural disasters, the homeland security scope is defined by the US government to consist of (Wikipedia, 2007):

-Domestic intelligence activities, largely today within the FBI;
-Critical infrastructure protection;
-Border security, including both land and maritime
-Transportation security, including aviation and maritime transportation
-Biodefense;
-Detection of nuclear and radiological materials.

### 4.3. Risk position

A simple model may be used to estimate a country risk position, as follows:

Risk position = $\alpha_1$ * SEI position + $\alpha_2$ * Unreadiness + $\alpha_3$ * Adverse exposure

SEI position varies from 1 to 5.

A score of 5 is only valid for a SB6 country.

> 1: Considered enemy by all SB6 countries; eg., Iran
> 2: Has no friends among the SB6 countries; Most African countries
> 3: Friendly developing countries; eg., Tunisia
> 4: Industrial countries not part of the SB6; eg. Japan
> 5: One of the SB6 countries

Unreadiness is assessed based on a country exposure to natural, technological, biological, or man-made threats. This may be measured in terms of a country weaknesses in terms of homeland security, business continuity, and disaster recovery.

Unreadiness scores are defined as follows:

> 1: Considered very weak in all of these: homeland security, business continuity, and disaster recovery;
> 2: Considered weak in all of these: homeland security, business continuity, and disaster recovery;
> 3: Considered weak in one of these: homeland security, business continuity, and disaster recovery;
> 4: Considered strong in homeland security, business continuity, and disaster recovery;
> 5: Considered very weak in all of these: homeland security, business continuity, and disaster recovery;

Adverse exposure is measured in terms of the peacefulness of the country. Scores may be defined as follows:

> 1: Has no enemies and is neutral; eg., Vatican, Switzerland
> 2: Has no enemies but not neutral; eg. Many developing countries.
> 3: No in any war; eg., Many developing countries.
> 4: In war but not considered a terrorist country; eg. USA, UK, Poland.
> 5: Considered a terrorist country by many countries; eg., North Korea, Iran, Syria.

We next propose some metrics that may be useful in estimating the country risk position and its priorities.

## 5. Security metric and prioritization

We propose a generic fusible scale that may be used to measure threats, business impact, and risks. The individual scale consists of 5 points defined as follows:

| Linguistic Term | Score |
|---|---|
| V: Very low; | 1 |
| L: Low; | 2 |
| M: Moderate; | 3 |
| H: High; | 4 |
| X: Very high | 5 |

V: Very low;
L: Low;
M: Moderate;
H: High;
X: Very high.

The fused scale is given by the following matrix:

| Linguistic Fused Scale | | | | | | |
|---|---|---|---|---|---|---|
| Matrix entries are the fused scale, eg, Risks | | | | | | |
| | | Scale 2:eg, Business Impact | | | | |
| | | V | L | M | H | X |
| **Scale 1: eg, Threats** | V | V | VL | L | LM | M |
| | L | VL | L | LM | M | MH |
| | M | L | LM | M | MH | H |
| | H | LM | M | MH | H | HX |
| | X | M | MH | H | HX | X |

| Quantitative Fused Scale | | | | | | |
|---|---|---|---|---|---|---|
| Matrix entries are the fused scale, eg, Risks | | | | | | |
| | | Scale 2:eg, Business Impact | | | | |
| | | V | L | M | H | X |
| **Scale 1: eg, Threats** | V | 1 | 1.5 | 2 | 2.5 | 3 |
| | L | 1.5 | 2 | 2.5 | 3 | 3.5 |
| | M | 2 | 2.5 | 3 | 3.5 | 4 |
| | H | 2.5 | 3 | 3.5 | 4 | 4.5 |
| | X | 3 | 3.5 | 4 | 4.5 | 5 |

The target assets candidate for protection may be prioritized as follows:

1. Estimate the weakness of any security controls or safeguards that are currently in use in the protection of the candidate asset, and assign it a point of the individual quantitative scale, from 1 to 5, given above.

2. Estimate risks for all candidate business assets, as shown in Tables 1 to 3, and obtain a number between 1 and 5.

3. Compute priority= risk * weakness

4. Sort candidate assets in descending order of priorities; and obtain a priority vector.

The priority vector will be need in a country definition of its security strategy given its risk position.

## 6. Conclusion and validation for the study

This work presents a pilot study that proposes a model that assembles decision support information needed for the development of a country's security strategy. The decisional support is given in terms of a country's risk position estimated in terms of the country's position on security economic intelligence, security unreadiness, and adverse exposure.

Interested scholars may take this work in progress a step further and build an experimental design for the purpose of testing the model.